\documentclass[11pt,twoside]{article}

\usepackage{asp2004}
\usepackage{epsf}
\usepackage{psfig}
\usepackage{lscape}

\begin{document}
\title{White dwarfs in the European Galactic Plane Surveys (EGAPS)}
%%% Fill in title
\author{L. Morales-Rueda,$^1$ P. J. Groot,$^1$ R. Napiwotzki,$^2$
  J. Drew,$^3$ and the EGAPS collaboration} %%% Fill in author names
  \affil{$^1$ Department of Astrophysics, Radboud Universiteit
  Nijmegen, PO Box 9010, Nijmegen 6500 GL, The Netherlands\\$^2$
  Centre for Astrophysics Research, STRI, University of Hertfordshire,
  Hatfield AL10 9AB, UK\\$^3$ Astrophysics Group, Department of
  Physics, Imperial College London, Exhibition Road, London SW7 2AZ,
  UK} %%% Fill in author affiliations

\begin{abstract} %%% Abstract to run on from here.
The space density of white dwarfs is highly uncertain even
nearby. This results from the fact that the known sample of white
dwarfs is largely incomplete in part because most white dwarfs have
been discovered as by-products in non-dedicated surveys. In order to
obtain more accurate white dwarf space densities and scale heights we
must build up a complete sample of white dwarfs. The European Galactic
Plane Surveys (EGAPS) are the best database to search for white dwarfs
as they will provide broad band (U, g', r', i') and narrow band
(H$\alpha$ and HeI) measurements for one per cent of all the stars in
the Galaxy. By looking at the Galactic Plane, where most stars are, we
ensure that we are obtaining a complete sample. The space densities
obtained from EGAPS can then be compared with those found in high
latitude surveys such as the Sloan Digital Sky Survey (SDSS). The
methods used to identify white dwarfs using the colours available in
EGAPS are described and some preliminary results presented.
\end{abstract}

\section{European Galactic Plane Surveys: EGAPS}

The space density of white dwarfs is not well known even in the solar
neighbourhood \citep{s04}. There is also a deficit of bright white
dwarfs in the Galactic Plane, compared to high Galactic latitudes, due
to the fact that most white dwarfs have been discovered out of the
Plane as by-products of extragalactic surveys. The best way to obtain
a complete sample of white dwarfs, and thus compute accurate space
densities, is to search for them in the Galactic plane, where most of
them reside. The use of a multi-band Galactic Plane Survey will
greatly facilitate this task.

EGAPS is the combination of a number of Galactic Plane surveys in
several passbands. The main aim of EGAPS is to obtain broad band (U,
g', r', i', Z, Y, J, H, Ks) and narrow band (H$\alpha$ and HeI)
photometry of a 10 degree latitude strip centred in the Plane all
along the Galaxy, i.e. covering the Northern and Southern Galactic
Planes, and going down to 21st magnitude in the optical bands. Most of
the stars in the Galaxy lie in the Plane thus by surveying a 10
degree-wide strip we are gathering photometry for 1 billion stars, one
per cent of the total number of stars in the Milky Way. These large
numbers are required to study statistically the different stellar
populations.

The surveys that make up EGAPS (on-going, approved and proposed) are
shown, together with their coverage, filters and limiting magnitudes,
in Table 1.

IPHAS, UVEX and VPHAS+ are all double pass surveys and include
re-observations at intervals of at least two years to determine proper
motions of all targets. $\Omega$White overlaps in area with VPHAS+ and
includes observations in the HeI narrow-band filter and 25
observations of selected fields in a two-hour interval. This will
allow the detection and study of HeI emission sources such as AM~CVn
systems and of short period variables such as AM CVns, magnetic
cataclysmic variables (CVs), post-bounce CVs and other ultracompact
binaries and fast pulsators.

\begin{table}[!ht]
\caption{IPHAS: INT/WFC Photometric H$\alpha$ Survey of the Northern
  Galactic Plane \citep{d05}. UVEX: Northern Galactic Plane UV-Excess
  Survey. VPHAS+: VST/OMEGACAM Photometric H$\alpha$ Survey of the
  Southern Galactic Plane. $\Omega$White: VST/OMEGAWHITE variability
  Survey. VVV: VISTA Variables in the V\'{\i}a L\'{a}ctea.}
\smallskip
\begin{center}
\begin{tabular}{lllll}
\tableline
\noalign{\smallskip}
Survey & Area (deg$^2$) &  Filters & Limit mag \\
%       &  (deg$^2$) &         & mag &             \\
\tableline
\noalign{\smallskip}
\multicolumn{4}{c}{Northern Hemisphere} \\
\noalign{\smallskip}
IPHAS & 10$\times$180 & H$\alpha$,r',i' & r'=21 \\
UVEX & 10$\times$180 & U,g',r',HeI & g'=22 \\
%\tableline
\multicolumn{4}{c}{Southern Hemisphere} \\
\noalign{\smallskip}
VPHAS+ & 10$\times$180 & u',g',r',i',H$\alpha$ & r'=21 \\
$\Omega$White & 400 & u',g',r',i',H$\alpha$,HeI & g'=22 \\
VVV & 520 & Z,Y,J,H,Ks\\
\noalign{\smallskip}
\tableline
\end{tabular}
\end{center}
\end{table}

The science drivers for EGAPS are very broad and range from the study
of proto-stars, interacting binaries, planetary nebula and stellar
remnants, such as white dwarfs, to the study of Galactic structure and
dust. IPHAS will most probably be complete by the end of 2007. The
completion date for UVEX is still uncertain. Both VST programs (VPHAS+
and $\Omega$White) will probably start at the end of 2007/beginning of
2008 and will be complete by 2010. The VVV survey has yet to be
approved. Up to date information on EGAPS and its constituent surveys
will be posted during and after the data taking at:
http://www.egaps.org

\section{Searching for white dwarfs in EGAPS}

White dwarfs are a galactic population and yet have mostly been found
at high latitudes. As mentioned above, the two main reasons for this
are: 1.  they have been found as interlopers in the search for blue
extragalactic sources, 2. the plane presents a challenge to observe
due to the vast number of sources and the high extinction. As most of
the white dwarfs in the Galaxy will be in the plane, it is definitely
worth looking for them there. From simulations of the distribution of
objects with Galactic latitude, using a galactic model based on
\citet{bp99} and including extinction towards the Galactic plane
according to the Sandage model, Nelemans (private communication) finds
that 12 (40) per cent of old (young) white dwarfs, i.e. with Mv = 15
(10), are concentrated within $|b|< 5\deg$. Reddening turns out not to
be such a tough problem to go around as it is only severe within the
first two degrees in latitude.

Synthetic spectra for 6 hydrogen-rich white dwarfs with temperatures
of 10,000, 20,000, 30,000, 40,000, 50,000 and 60,000 K are shown in
Fig.~1 together with the passbands of broad band filters used in EGAPS
and the passband of the H$\alpha$ narrow band filter. The H$\alpha$
narrow band filter allows us to search for both H$\alpha$ emission and
absorption stars, this last group includes white dwarfs. The H$\alpha$
equivalent width of hydrogen-rich white dwarfs scales with
temperature, being larger for temperatures between 10,000 and 20,000 K
(see Fig.~2) which means that the IPHAS survey on its own is most
sensitive to white dwarfs in that temperature range. This temperature
range is particularly interesting as it overlaps with the
hydrogen-rich white dwarf instability strip, 11,000--12,500 K
\citep{m06}.

\begin{figure}[!ht]
\plotfiddle{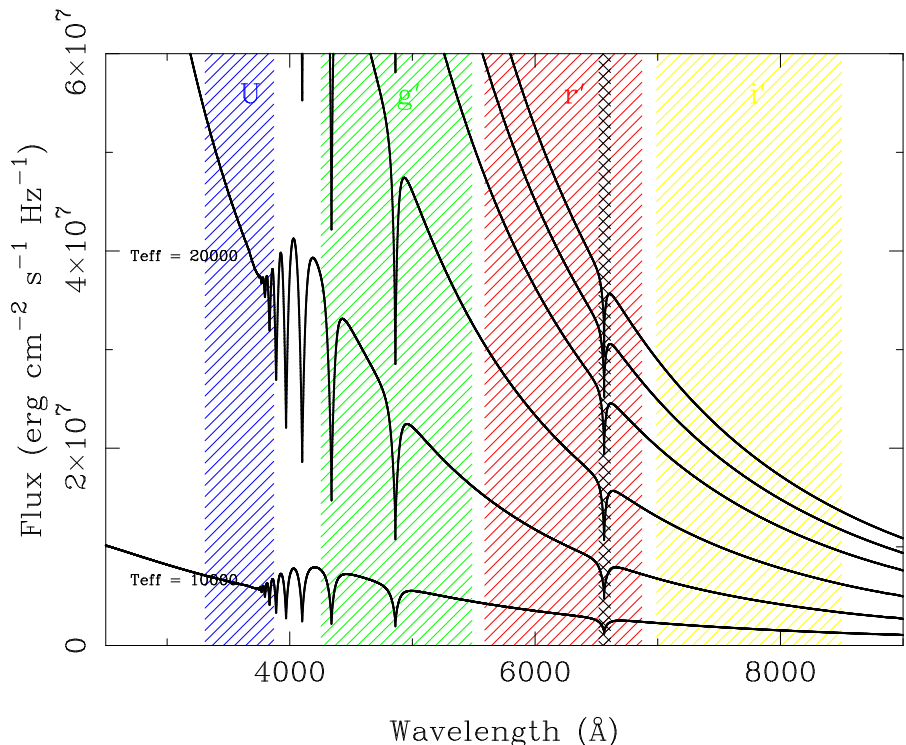}{8.8cm}{-90}{53}{53}{-210}{295}
\caption{Synthetic spectra of six hydrogen-rich white dwarfs of
  different temperatures (10,000 to 60,000 K in 10,000 K steps)
  obtained with TLUSTY. Note that the flux increases with
  temperature. Also plotted (hatched) are the bandpasses of the broad
  band filters U, g', r', and i', and the the bandpass of the narrow
  band filter H$\alpha$ (cross-hatched).}
\end{figure}

\begin{figure}[!ht]
\plotfiddle{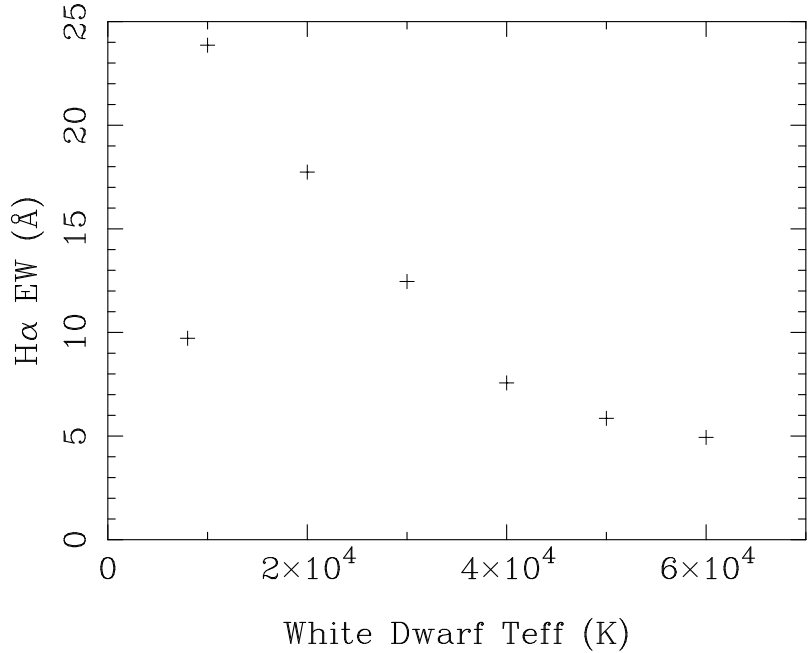}{8.5cm}{-90}{58}{58}{-220}{310}
\caption{Equivalent width (EW) of the H$\alpha$ absorption line for
  hydrogen-rich white dwarfs of different temperatures. Note that the
  EW is larger for white dwarf with temperatures between 10,000 and
  20,000 K.}
\end{figure}

\subsection{The IPHAS colour-colour space}

Fig.~3 shows an example of the IPHAS colour space for a given field in
Cassiopeia. The position where the un-reddened hydrogen-rich white
dwarfs lie is marked with a box. The position of this box was
calculated by convolving the IPHAS band passes with a set of SDSS
hydrogen-rich white dwarfs. For this given field, we find one white
dwarf candidate. The rest of the stars in the field outline the main
sequence, as well as the giant and supergiant sequences (see
\citeauthor{d05} 2005, for a description of the positions of the
un-reddened and reddened main sequence, the giant and the supergiant
sequences in the IPHAS colour space).

White dwarfs in the IPHAS colour space lie away from the main sequence
and therefore are easy to pick out.

\begin{figure}[!ht]
\plotfiddle{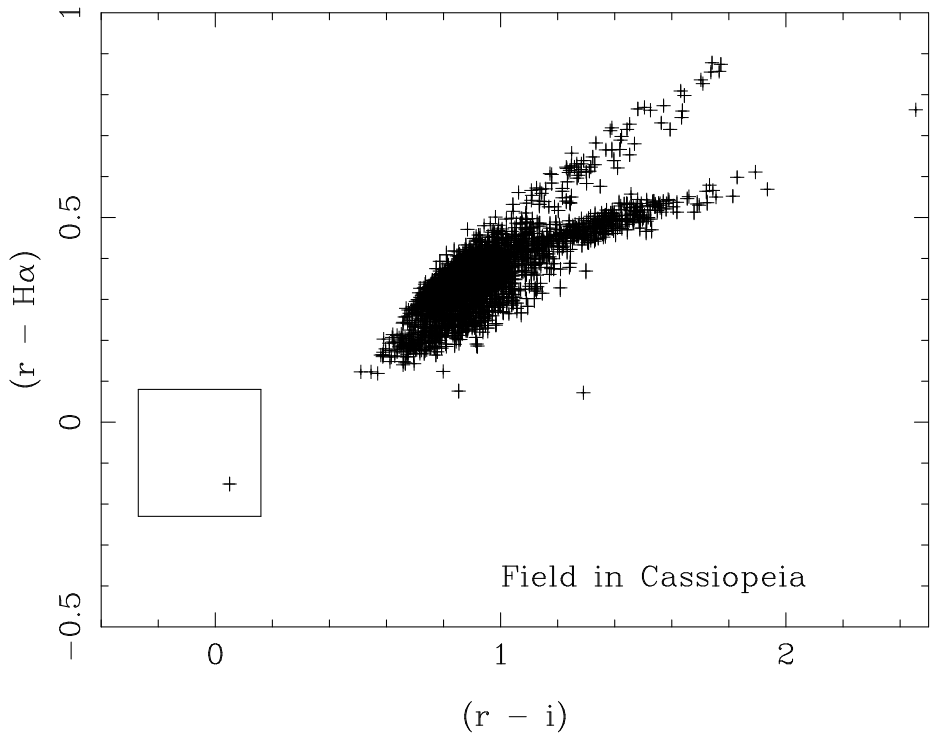}{8.5cm}{-90}{55}{55}{-220}{310}
\caption{IPHAS colour space indicating the position of hydrogen-rich
  white dwarfs with respect to the main sequence and the giant and
  supergiant sequences.}
\end{figure}

\section{Preliminary Results}

We performed a search for white dwarfs by looking for candidates in
the colour box presented in the previous section in the first IPHAS
data release, which comprises 10 observing epochs between August 2003
and 2004. As IPHAS is a double pass survey, all fields are observed
twice with only a small region of non-overlap between the two
fields. We only selected candidates that appeared in both fields and
whose positions differed, from one field to its overlap, by one arcsec
at the most.

The densities of white dwarfs found vary considerably from epoch to
epoch and are given in Table 2. These discrepancies cannot be
explained only by the different area in the sky that each epoch
covers, or by contamination of the white dwarf box due to different
reddening in different fields. The result of reddening would be to
move white dwarfs to larger values of (r $-$ i) and (r $-$ H$\alpha$)
out of the box, thus the white dwarf density would decrease. No other
objects would fall in the white dwarf box to increase the number of
false detections so this cannot explain the large value seen, for
example in Jun 2004. The most probable causes for the discrepancies
are: the fact that not all the data is of the same quality and only a
seeing cutoff has been applied in this search, and that there is not,
as yet, a global photometric solution for the survey, which translates
in offsets between colours for different fields.

\begin{table}[!ht]
\caption{Density of white dwarf candidates found for the first 10
  epochs of observations of IPHAS.}
\smallskip
\begin{center}
\begin{tabular}{llllll}
\tableline
\noalign{\smallskip}
Epoch & Area & Density &  Epoch & Area & Density\\
 & (deg$^{2}$) & (deg$^{-2}$) & & (deg$^{2}$) & (deg$^{-2}$)\\
\tableline
\noalign{\smallskip}
Aug 2003 & 72 & 0.26 & Sep 2003 & 14 & 4.6\\
Oct 2003 & 145 & 1.6 & Nov 2003 & 199 & 0.6\\
Dec 2003 & 118 & 0.8 & Jun 2004 & 96 & 12.2\\
Jul 2004a & 71 & 4.6 & Jul 2004b & 99 & 2.2\\
Aug 2004a & 58 & 0.76 & Aug 2004b & 146 & 2.18\\
\noalign{\smallskip}
\tableline
\end{tabular}
\end{center}
\end{table}

If we only make use of the data presented by \cite{d05} (7 fields,
1.86 deg$^2$), which has been checked for quality and calibration, we
find 1 un-reddened hydrogen-rich white dwarf, giving a density of 0.6
deg$^{-2}$. The theoretical prediction is of 0.8 white dwarfs per
deg$^2$ (assuming all spectral types and temperatures). From follow up
spectroscopy using the multi-object spectrograph HECTOSPEC mounted on
the 6.5\,m MMT (Steeghs et al in preparation), we also find an average
of 1 hydrogen-rich white dwarf per deg$^2$.

Preliminary follow-up of UVEX sources shows them to be white dwarfs at
a density of 1-2 per pointing (4-8 per deg$^2$).

\section{Conclusions}

EGAPS will constitute the perfect collection of surveys to obtain a
complete sample of white dwarfs, determine their space densities and
their scale height. On their own, some of the surveys that make up
EGAPS, e.g. IPHAS, will also provide the data to find white dwarfs,
targeting systems in specific temperature ranges. The temperature
range that IPHAS is sensitive to includes the instability strip
allowing us to find pulsating white dwarfs in large numbers. Following
up of these objects can lead to asteroseismological studies to
determine their internal structure and to search for planets around
them.

\acknowledgements This paper makes use of data from the Isaac Newton
Telescope, operated in the island of La Palma by the ING in the
Spanish Observatorio del Roque de los Muchachos of the Instituto de
Astrof\'{\i}sica de Canarias. LMR and PJG are funded by NWO-VIDI grant
39.042.201 to PJG. LMR would like to thank the Leids Kerkhoven-Bosscha
Fonds for funding her attendance to this workshop.

\end{document}